\documentclass{article} 
\usepackage[final]{colm2026_conference}

\usepackage{microtype}
\usepackage{hyperref}
\usepackage{url}
\usepackage{booktabs}
\usepackage{subcaption}
\usepackage{graphicx}
\usepackage{wrapfig}
\usepackage[pass]{geometry}

\usepackage{booktabs}
\usepackage{graphicx}
\usepackage{wrapfig}
\usepackage{caption}
\usepackage{tabularx}
\usepackage{amsmath}
\usepackage{hyperref}
\usepackage{multirow}
\usepackage{makecell}
\usepackage{xcolor}
\usepackage{enumitem}

\usepackage{lineno}

\definecolor{darkblue}{rgb}{0, 0, 0.5}
\hypersetup{colorlinks=true, citecolor=darkblue, linkcolor=darkblue, urlcolor=darkblue}

\title{Towards Comprehensive Mobile Application Testing for User-centric Feature Coverage via Simulating Real-World Users with Personas}

\author{
  Ha Min Son\textsuperscript{1,2} \And
  Huan Ren\textsuperscript{2} \And
  Huan Song\textsuperscript{2} \And
  Zhe Zhao\textsuperscript{1} \And
  Xin Liu\textsuperscript{1} 
  \AND
  \begin{tabular}{@{}l@{}}
    \textsuperscript{1}University of California, Davis \\
    \textsuperscript{2}CodeDroid LLC
  \end{tabular}
}

\setlist{nosep, leftmargin=*}

\begin{document}

\ifcolmsubmission
\linenumbers
\fi

\maketitle

\begin{abstract}
Large Language Model (LLM) agents are increasingly used for black-box mobile application testing, where the ultimate goal is the comprehensive coverage of the features users rely on. However, most existing studies focus on Activity Coverage, which measures the proportion of distinct application screens visited during a test session but does not directly measure whether agents actually test the features users care about. To address this, we focus on two research questions: 
(1) how to automatically measure whether a testing framework covers the features users care about, and (2) how to design a framework that achieves more comprehensive feature coverage. To answer (1), we introduce the User-Centric Feature Coverage benchmark, which consists of two components: a manually crafted dataset of 3,012 user-centric target features based on real reviews and app descriptions, and an automated evaluation pipeline that translates agent-generated test cases and target features into a shared semantic space, then uses an LLM-as-a-judge to compute coverage. To answer (2), we introduce ECS (\textbf{E}xploration, \textbf{C}ritical User Journeys, and \textbf{S}ynthetic Reviews). ECS is a framework that explores the application, then uses an LLM to convert exploration traces into Critical User Journeys (CUJs) to bring the semantic meaning of the trace closer to a distinct feature. Each CUJ consists of an LLM-inferred user intent paired with reproducible test steps. ECS then aggregates multiple CUJs using persona-augmented synthetic reviews to capture more complex features, simulating real users who build knowledge of an application across multiple screens before writing a review. ECS achieves 64.1\% feature coverage, a 23.8\% gain over the state-of-the-art baseline, and covers 72.5\% of features in app descriptions, capturing the most prominent features. Our failure analysis shows statelessness and exploration discoverability as main reasons of failures, providing future directions.


\end{abstract}
 
\begin{wrapfigure}{r}{0.3\textwidth}
    \centering
    \vspace{-20pt}
    \includegraphics[width=\linewidth]{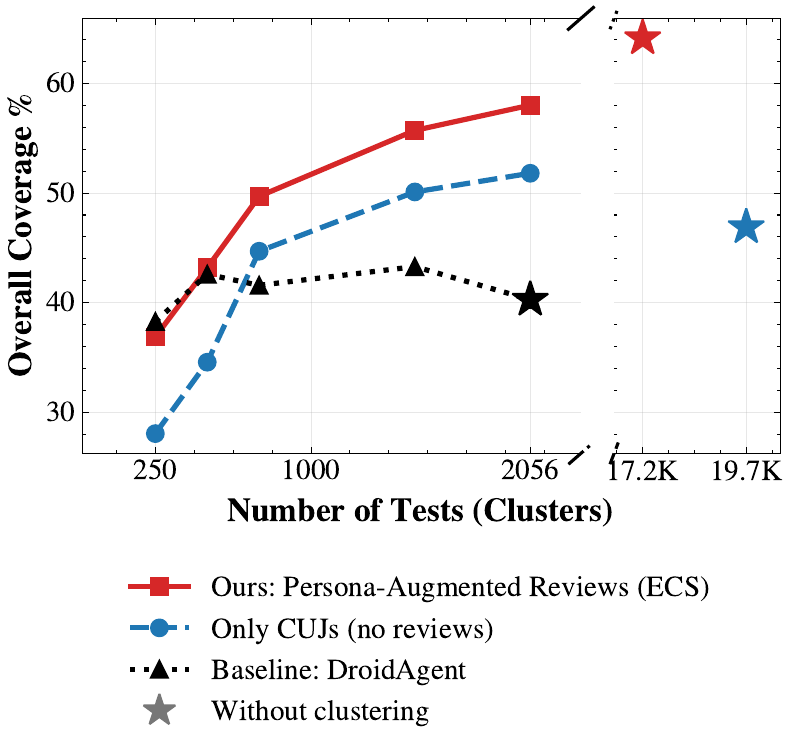}
    \caption{Total target feature coverage (3012 features).}
    \label{fig:main_coverage}
    \vspace{-10pt}
\end{wrapfigure}

\section{Introduction}
Black-box mobile application testing using LLM agents has shown significant progress~\citep{liu2023chatting,yoon2023autonomous,wang2025llmdroid}. In this setting, the input is the application binary (APK), with no access to source code, and the output is a state graph of the app, natural language test cases, or exploration traces. However, most existing studies focus on Activity Coverage and have limitations. Activity Coverage measures the proportion of distinct application screens, called activities, visited during a test session. While ground truth activity counts are obtained from the APK, making the metric well-defined and reproducible, Activity Coverage does not directly measure whether agents test the features users actually care about, which is the ultimate goal of testing. 
In fact, developer surveys indicate that developers often prefer use-case and functional tests over structural metrics such as Activity Coverage~\citep{linares2017developers,yoon2023autonomous}. Here, we define a \textbf{feature} as a discrete application capability for accomplishing an objective, which often spans a sequence of activities and operates at a higher semantic level than activity visits.
A \textbf{user-centric} feature grounds features in real-world human expressions. For example, an app description captures features intended by developers (e.g., "Price drop alerts for saved items"), a positive review typically mentions a successful feature (e.g., "Check boxes in the cart are a good idea"), and a negative review often shows failures or edge cases (e.g., "Shopping cart silently drops items").


Despite the importance of testing for feature coverage, there is a notable gap in the literature for automatic and scalable evaluation. This is largely because user-centric features are difficult to define and lack established ground truth datasets.
As a notable exception, DroidAgent~\citep{yoon2023autonomous} introduced functional feature coverage using manual human counting. This was a conceptually important contribution, but has two limitations: (1) manual counting does not scale across many baselines and applications, nor does it provide a list of features that were not captured, and (2) the agent ultimately achieves low overall feature coverage (Figure~\ref{fig:main_coverage}). This motivates our work and drives two research questions.

\begin{figure*}[t]
    \centering
    \includegraphics[width=0.8\textwidth]{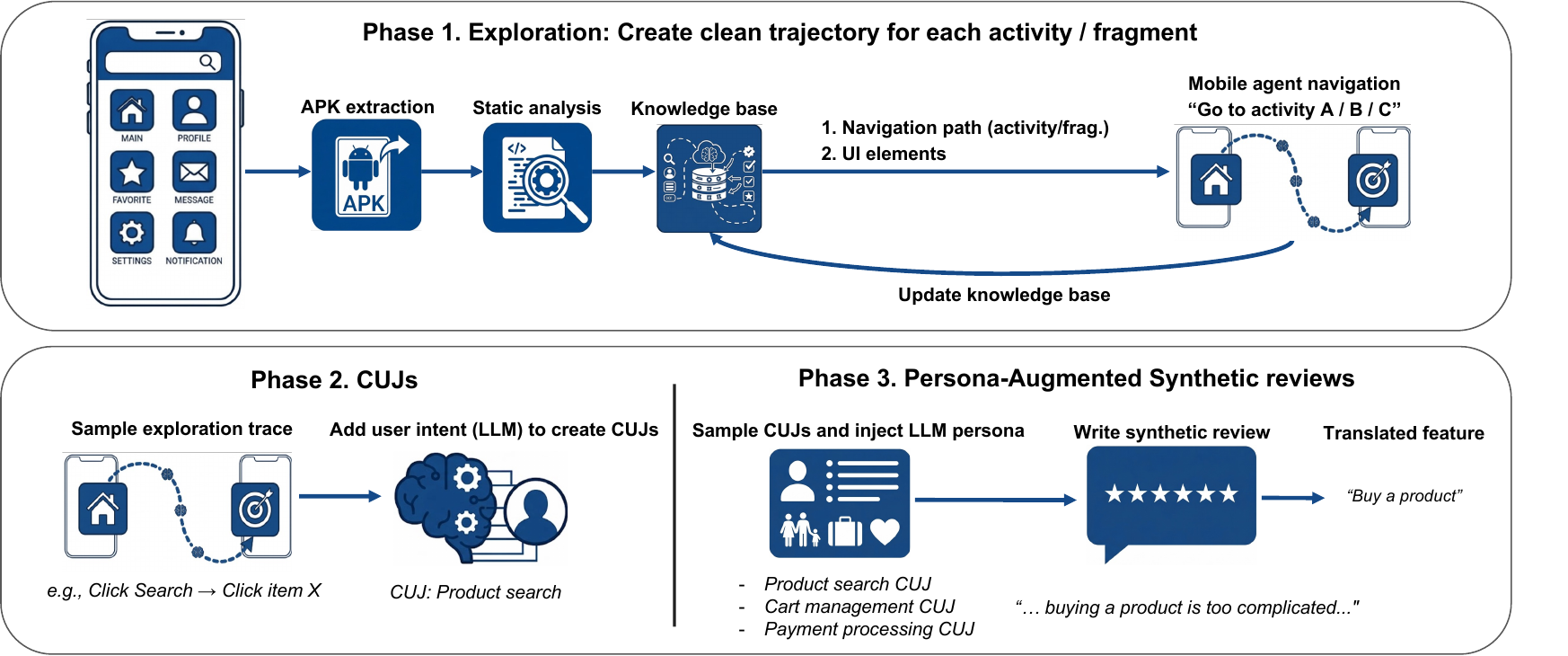}
    \caption{Overview of the ECS pipeline. Structural app exploration traces are mapped into intent-annotated Critical User Journeys (CUJs); batches of CUJs fed to a persona-augmented LLM to generate synthetic reviews.}
    \label{fig:method_overview}
    \vspace{-10pt}
\end{figure*}

\textbf{RQ1: Can we directly and automatically measure user-centric feature coverage?}
We introduce the User-Centric Feature Coverage benchmark, which consists of two components. The first is a dataset of 3,012 target features gathered from Play Store app descriptions and user reviews across 10 popular applications. While the ideal ground truth would be internal requirements documents or formal feature specifications, these are rarely public. We instead rely on app descriptions and user reviews as a proxy for ground truth features~\citep{motger2024t,maalej2015bug,gorla2014checking}. The second is an automated evaluation pipeline that translates agent-generated test cases and target features into a shared semantic space, then uses an LLM-as-a-judge to compute coverage without human annotation. To validate the pipeline, we manually judged 100 randomly sampled feature pairs and found 93\% agreement with the LLM judge (Cohen's $\kappa$ = 0.83) suggesting strong reliability. However, the state-of-the-art, DroidAgent, achieves low feature coverage (Figure~\ref{fig:main_coverage}), motivating our next question.

\textbf{RQ2: Can we design a testing framework that achieves more comprehensive feature coverage?}
We introduce ECS (\textbf{E}xploration, \textbf{C}ritical User Journeys, and \textbf{S}ynthetic Reviews), a framework that explores the application, then uses an LLM to convert the exploration traces into Critical User Journeys (CUJs).
By combining traces with LLM-inferred user intent and reproducible test steps, each CUJ brings the semantic meaning of the trace closer to a feature. ECS then uses persona-augmented synthetic reviews to combine disjointed CUJs and capture more complex features, simulating real users who build holistic knowledge of the application before writing a review. Personas are generated from reviews of entirely different applications, and exploration traces are generated independently of the target feature list (Figure~\ref{fig:method_overview}).
As shown in Figure~\ref{fig:main_coverage}, ECS achieves 64.1\% feature coverage, a 23.8\% gain over DroidAgent, and a 17.2\% improvement over our method without synthetic reviews. In addition, we cover 72.5\% of the app description, which likely represent the most prominent features. ECS also achieves state-of-the-art Activity Coverage on both small traditional applications and large commercial applications (Table~\ref{tab:exploration}). Our ablations (Sec.~\ref{sec:ablations}) show that persona diversity improves coverage while fixed emotional personas cause degradation. Analysis of the missed features shows that LLMs consistently fail on stateful behaviors requiring cross-session reasoning, as well as features constrained by exploration discoverability. Table~\ref{tab:vocabulary} defines the vocabulary used throughout this paper.

\begin{table}[ht]
\centering
\vspace{-0pt}
\caption{Core Vocabulary and Running Example}
\label{tab:vocabulary}
\scriptsize 
\setlength{\tabcolsep}{4pt} 
\begin{tabular}{p{0.18\linewidth}p{0.35\linewidth}p{0.37\linewidth}}
\toprule
\textbf{Term} & \textbf{Definition} & \textbf{Example} \\
\midrule
\textbf{Exploration Trace} & A raw chronological log of system interactions and visible screens/UI elements (all in text form). & Clicking an ``Add to Cart'' button in ``Search'', followed by a screen transition to the ``Shopping Bag'' activity. \\
\addlinespace
\textbf{Critical User Journey (CUJ)} & An exploration trace paired with an LLM-inferred user intent and explicit test steps. & ``Quick Add to Cart from search results. Steps: 1. Tap item ``X''. 2. Tap Add to Cart.'' \\
\addlinespace
\textbf{Synthetic Review} & An LLM-generated narrative that combines multiple CUJs to simulate a real user experience. & ``The 'Add to Cart' button directly in the search results is a coin flip... No confirmation, no update to the Cart tab.'' \\
\addlinespace
\textbf{Real Review} & An authentic user comment from the Play Store. & ``Cannot add items to the cart.'' \\
\addlinespace
\textbf{Translated Feature} & A normalized, objective description of a capability, stripped of emotional language (by an LLM). & ``Add-to-cart functionality.'' \\
\addlinespace
\textbf{Feature Coverage} & An LLM judge evaluating whether the translated features extracted from synthetic reviews and CUJs cover the app descriptions and real user reviews (Pass/Fail). & \textit{Syn.}: ``Add to Cart from search results'' \newline \textit{Real}: ``Add-to-cart functionality.'' \newline \textit{Judge}: \textbf{Pass.} \\
\bottomrule
\end{tabular}
\vspace{0pt}
\end{table}

Our contributions are summarized as follows:
\begin{enumerate}[leftmargin=*, label=(\arabic*)]
    \item \textbf{A User-Centric Feature Coverage Dataset}: We publicly release a dataset of 3,012 target features crawled from 5,000 real user reviews and app descriptions across 10 popular applications, capturing user-centric features which Activity Coverage does not directly measure.
 
    \item \textbf{An Automated Evaluation Pipeline}: Using this dataset, we introduce an automated LLM-as-a-judge evaluation pipeline that translates agent-generated test cases and target features into a shared semantic space to compute user-centric feature coverage without human annotation. Together, the dataset and pipeline create the User-Centric Feature Coverage benchmark.
    
    \item \textbf{The ECS Testing Framework}: 
    We propose ECS, a framework that explores the application, then uses an LLM to convert exploration traces into Critical User Journeys (CUJs), and aggregates those CUJs using diverse, persona-augmented synthetic reviews to capture complex features, simulating real users who build holistic knowledge of an application before writing a review. ECS achieves 64.1\% feature coverage, a 23.8\% gain over the state-of-the-art baseline, and covers 72.5\% of target app descriptions, capturing the most prominent features.

    \item \textbf{Failure analysis}:  We analyze and categorize the 1,081 features our method failed to capture, such as statelessness and discoverability. Specifically, agents struggle to capture features requiring persistent memory across app restarts, and they frequently miss multi-step, cross-screen interactions. This indicates critical future directions.
\end{enumerate}

\vspace{-5pt}
\section{Related Work}
\vspace{-5pt}
 
\textbf{Automated Mobile App Testing.}
Early automated testing tools used random event generation~\citep{developers2017ui} or model-based exploration~\citep{li2017droidbot}. Deep learning and reinforcement learning approaches subsequently improved coverage. Humanoid~\citep{li2019humanoid} trained a deep neural network on human click traces, FastBot~\citep{lv2022fastbot2} used a Q-table to reward discovery of new activities, and DinoDroid~\citep{zhao2024dinodroid} extended this with a deep Q-network. 
More recently, LLM-based agents achieved high Activity Coverage by instructing LLMs to select actions based on UI trees and task descriptions. GPTDroid~\citep{liu2023chatting} provides the LLM with a list of activities and a summary of prior actions, and LLMDroid~\citep{wang2025llmdroid} combines both paradigms, using RL-based exploration and switching to an LLM when coverage plateaus.  Lastly, DroidAgent~\citep{yoon2023autonomous} asks the LLM to first infer a test case targeting an app feature, then navigate to new activities to execute it. The agent is provided with a list of all known activities to guide this process. However, this design has two limitations. First, because the agent must infer a test case before navigating, there is no guarantee it will attempt to visit any particular activity. Second, inferring features for a screen it has never visited makes it unlikely to find less prominent features, since the LLM has no grounding in the actual UI elements present. Static analysis was already used to guide exploration as early as 2013~\citep{azim2013targeted}, but most LLM-based approaches treat the LLM as a simulation of a real user and rely on exploration alone. We use static analysis to provide the LLM with a richer starting point, since LLMs, unlike real users, are able to digest this large amount of information.

\textbf{Feature-Level and Use-Case Testing.}
Developer surveys showed that developers often prefer use-case completeness over structural metrics such as Activity Coverage~\citep{linares2017developers}. DroidAgent~\citep{yoon2023autonomous} introduced Feature Coverage using human annotation, which was a conceptually important contribution but is limited by the absence of a target feature list and by the cost of manual counting. We extend the feature coverage concept to be user-centric and introduce a fully automated pipeline with an independent, a priori proxy dataset of features gathered from publicly available data.

\textbf{Mining the Play Store and LLMs as Evaluators.}
Prior work validates user reviews and app descriptions as reliable proxies for user-centric features~\citep{motger2024t,maalej2015bug,gorla2014checking}, and LLM-as-a-Judge~\citep{zheng2023judging} has become a standard evaluation paradigm, though known risks include self-preference bias~\citep{wang2024large}. We discuss both lines of work further in Appendix~\ref{appendix:related_extended}.

 

 \vspace{-5pt} 
\section{Building the User-Centric Feature Coverage Benchmark}
 \vspace{-5pt}

We introduce the User-Centric Feature Coverage benchmark with two components: a dataset of 3,012 target features across 10 popular applications, and an automated evaluation pipeline for measuring how well a testing framework covers them.

\begin{wraptable}{r}{0.5\textwidth}
\vspace{-13pt}
\centering
\caption{Filtering taxonomy for review pre-processing, ordered by occurrence frequency.}
\label{tab:filter_taxonomy}
\scriptsize
\resizebox{\linewidth}{!}{
\begin{tabular}{lrl}
\toprule
\textbf{Category} & \textbf{Count} & \textbf{Representative Example} \\
\midrule
Subjective \& Vague & 491 & ``App has become a convoluted mess.'' \\
Billing \& Pricing & 345 & ``Charged after attempting to cancel.'' \\
Content  & 268 & ``File uploads get stuck as `Paused'.'' \\
Logistics \& Physical & 251 & ``Packages delivered to the wrong unit.'' \\
Policy \& Account & 190 & ``Not HIPAA compliant.'' \\
Customer Support & 167 & ``No human tech support available.'' \\
Cross-Platform & 93 & ``Unusable on Chromebook.'' \\
\bottomrule
\end{tabular}
}
\vspace{-10pt}
\end{wraptable}

\textbf{Building the target feature list.} We note that the ideal ground truth for a feature coverage benchmark would be internal requirements documents or formal feature specifications. Such documents are almost never publicly available, and to our knowledge no existing benchmark provides a realistic, large-scale feature list for popular commercial applications. We instead construct one from publicly available sources that prior work has validated as reliable proxies~\citep{motger2024t}. First, Play Store app descriptions capture the features developers deliberately built and chose to advertise~\citep{gorla2014checking,shamsujjoha2021checking}. In addition, user reviews capture realized utility~\citep{maalej2015bug,pagano2013user}. Because users are explaining why they continue to use, or have stopped using, an application, they have an incentive to mention less prominent features and even highlight \textit{missing features} (Appendix~\ref{appendix:missing}) that structural metrics do not directly measure. Review-derived feature sets are also more comprehensive than static, crowd-sourced lists because reviews are continuously updated~\citep{motger2024t}.

 \begin{wraptable}{r}{0.45\textwidth}
\vspace{-10pt}
\centering
\caption{Dataset before and after filtering. A review can mention multiple features, so totals exceed the number of crawled reviews.}
\label{tab:filtering_counts}
\scriptsize
\setlength{\tabcolsep}{4pt}
\begin{tabular}{lrr}
\toprule
\textbf{Source} & \textbf{Before} & \textbf{After} \\
\midrule
App Descriptions & 152 (2.6\%)  & 131 (4.3\%)  \\
Positive Reviews & 1220 (21.1\%) & 333 (11.1\%) \\
Negative Reviews & 4407 (76.3\%) & 2548 (84.6\%) \\
\midrule
\textbf{Total}   & \textbf{5779} & \textbf{3012} \\
\bottomrule
\end{tabular}
\vspace{-10pt}
\end{wraptable}

We collected data from 10 popular Google Play Store applications. For each, we extracted the full app description and 500 user reviews (250 most recent, 250 most relevant). Because individual reviews sometimes mention multiple distinct features, the raw text yields 5,779 feature candidates before filtering. However, the raw data contained substantial noise. Complaints about pricing, ads, and shipping logistics were common. None of these are software behaviors that an automated agent can observe. To eliminate them, we applied manual filtering using the taxonomy shown in Table~\ref{tab:filter_taxonomy}, reducing 5,779 feature candidates to a set of 3,012 target features (Table~\ref{tab:filtering_counts}).
These target features allow us to determine which features an agent found and which it missed.

\textbf{Creating an automated evaluation pipeline.}
LLM-generated synthetic reviews and target features are semantically different (Table~\ref{tab:vocabulary}). A synthetic review could say ``The `Add to Cart’ button directly in the search results is a coin flip\ldots,'' while the corresponding benchmark feature states ``Add-to-cart functionality.'' Direct string matching or embedding similarity is likely not reliable enough for a binary coverage decision. We address this with a two-step pipeline.

First, we prompt a separate language model, Gemini 3 Flash (temperature 0.0), to translate both agent outputs and target features into objective, stripped-down descriptions of the feature being tested, removing subjective language, emotional framing, and bug reports. We call the result a Translated Feature. Note that stripping emotions keeps the underlying feature. For example, the negative review ``Cannot add items to the cart'' is translated to ``Add-to-cart functionality'' rather than being discarded (Table~\ref{tab:vocabulary}). Translated features are then embedded using \texttt{gemini-embedding-001} into a 3,072-dimensional dense vector space.
Second, for each target feature we retrieve the top-5 candidate synthetic review or CUJ using cosine similarity, then ask an LLM-as-a-judge to make a binary coverage decision for each candidate pair. We retrieve 5 candidates to keep the candidate set small enough that the judge can evaluate each pair reliably in isolation. This pipeline is fully automated and can be applied to any agent that produces natural language test descriptions. We treat the resulting coverage score as our primary evaluation metric. To validate this approach, we manually judged 100 randomly sampled feature pairs and found 93\% agreement with the LLM judge (Cohen's $\kappa$ = 0.83), suggesting strong reliability.

Together, the target feature list and automated evaluation pipeline answer RQ1: feature coverage can be measured directly and automatically, without human annotation, by translating agent outputs and target features into a shared semantic space and using an LLM-as-a-judge for binary coverage decisions. 


\vspace{-5pt}
\section{ECS: Exploration, Critical User Journeys, and Synthetic Reviews}
\vspace{-5pt}

As shown in Figure~\ref{fig:method_overview}, our method, ECS, consists of three phases to generate comprehensive and reproducible test cases, and improve user-centric feature coverage: collecting exploration traces, mapping traces into Critical User Journeys (CUJs), and aggregating batches of CUJs into persona-augmented synthetic reviews.

\textbf{Phase 1: Exploration Traces.}
To effectively capture dynamic application environments for downstream LLM reasoning, we record chronological action trajectories by combining dynamic exploration with static code analysis. Using our e-commerce example (Table~\ref{tab:vocabulary}), an exploration trace would record the exact sequence of UI states and interactions, such as discovering a product screen and clicking an "Add to Cart" button. To optimize this exploration, we use the application's binary (the APK) and conduct static analysis on manifest declarations, resource identifier strings, and Intent filter definitions, which define how information is passed between activities (screens). All information about any UI elements or transition paths between activities is logged in a knowledge base. This provides the agent with knowledge of the app even before beginning exploration. The agent is then directed to navigate to each known activity in turn, and within each activity to find any available sub-screens (called \textit{fragments}). After each session, the knowledge base is updated with the paths and UI elements observed, allowing the agent to navigate further in subsequent sessions. Details of the agent are provided in Appendix~\ref{appendix:agent_actions}.
 
\textbf{Phase 2: Critical User Journeys.}
For each exploration trace, an LLM reads the sequence of visible UI elements, and infers two things: which features a user could have realized, and which expected features were absent from the observed state. A \texttt{missing\_gap} CUJ only counts toward coverage when it matches a real user-reported target. An imagined feature that no real user ever reported has nothing to match against, so it is never judged as covered (Appendix~\ref{appendix:isolation}). To prevent hallucination, all inferred features must be anchored to elements present in the visible text-based UI tree. Examples of the resulting reproducible test cases are shown in Appendix~\ref{appendix:cuj_examples}, and examples of the missing features successfully identified by the system are detailed in Appendix~\ref{appendix:missing}. We deliberately instruct the LLM to generate an inferred user intent for every interactable UI element on the screen. This exhaustive approach creates potential test cases for every interactable UI element, at the cost of generating highly redundant descriptions. The resulting output is a Critical User Journey (CUJ): a record of LLM-inferred user intent paired with the exact interaction steps observed during exploration. Because each CUJ includes explicit interaction steps, it functionally serves as a reproducible test case. 

\textbf{Phase 3: Simulating Real Users via Persona-Augmented Synthetic Reviews.}
A single CUJ captures only an isolated perspective from one exploration trace. However, it is likely that many complex features can only be captured by combining multiple CUJs. For example, buying a product could require a product search CUJ, a cart management CUJ, and a payment processing CUJ working together. To capture these complex features, we sample random batches of CUJs from across the application and prompt an LLM to write a synthetic review for each batch using a dynamically sampled user persona. The synthetic review is an aggregation method for combining disjointed CUJs, simulating real users who experience multiple CUJs, build holistic knowledge of the application, and then write reviews detailing the features they used.

Personas are constructed from real reviews of entirely different applications, preventing any data leakage from the benchmark. An LLM is prompted to create 40 to 50 user personas based on about 250 reviews from a random different app. Further discussion on the use of personas in language models is provided in Appendix~\ref{appendix:persona_related_work}. Our ablations in Section~\ref{sec:ablations} show that persona diversity is important. Fixed emotional personas such as a consistently angry or consistently lenient user degrade coverage, while our randomly sampled personas improve it.

Finally, to create a high-quality, reproducible test suite, and to control for the trade-off between exhaustively running every generated test and selecting a smaller, representative subset, we generate translated features for every CUJ and synthetic review. These translated features are then embedded using \texttt{gemini-embedding-001} into a 3,072-dimensional dense vector space, where we apply Agglomerative Clustering based on a target number of clusters to select a representative medoid text from each. The final outputs of ECS are the specific features covered by each test, accompanied by explicit test steps from individual CUJs or the batch of CUJs combined during synthetic review generation.

Together, the three phases of ECS answer RQ2: by combining structural exploration with persona-augmented synthetic review generation, our testing framework can achieve substantially more comprehensive feature coverage than methods optimizing for Activity Coverage alone (Figure~\ref{fig:main_coverage}).

\vspace{-5pt}
\section{Experimental Setup}
\vspace{-5pt}
We evaluate on 20 applications in two tiers. The first consists of 10 smaller open-source applications from the Themis benchmark~\citep{su2021benchmarking}, which have been widely used to evaluate activity coverage baselines. These apps lack ground truth feature sets and online reviews, so we use them only for Activity Coverage. The second consists of 10 top-ranking commercial applications from the Google Play Store, representing the scale and complexity of real-world deployment targets. These are the same apps used to build our benchmark. Full app lists are provided in Appendix~\ref{appendix:app_list}. For all baselines, each application is given a budget of 12 hours or 5,000 interaction steps, whichever is reached first. We use Gemini~3~Flash (\texttt{gemini-3-flash-preview}) for all LLM calls and \texttt{gemini-embedding-001} for all text embeddings. The prompts used for exploration, test generation, and evaluation are provided in Appendix~\ref{appendix:prompts}.

For exploration, we evaluate against three baselines DroidBot~\citep{li2017droidbot}, GPTDroid~\citep{liu2023chatting}, and DroidAgent~\citep{yoon2023autonomous}. For ECS, we force the agent to visit each target activity at least once within our budget. For each target activity, we give the agent 15 steps to explore before moving on. This forced exploration is useful to prevent the agent from getting stuck and improves discovery.

For feature coverage, DroidAgent is our main baseline because it is the only method that directly generates natural language test cases. We additionally adapt GPTDroid and LLMDroid by translating their step-level outputs into user goals before passing them to the judge (Appendix~\ref{appendix:baseline_adapt}). We test feature coverage under the same budget and against the same 3,012 target features. To generate our test cases, we task the LLM to generate CUJs of all interactable UI elements for each exploration trace, then we sample 16 CUJs 400 times with replacement to generate synthetic reviews. We include experiments with $K=2056$ clusters to match the number of test generated by DroidAgent. Coverage is then evaluated by retrieving the top-5 candidate synthetic reviews or CUJs for each target feature using cosine similarity and asking the LLM judge for a binary coverage decision.

\begin{figure*}[t]
    \centering
    \begin{subfigure}{0.27\textwidth}
        \includegraphics[width=\linewidth]{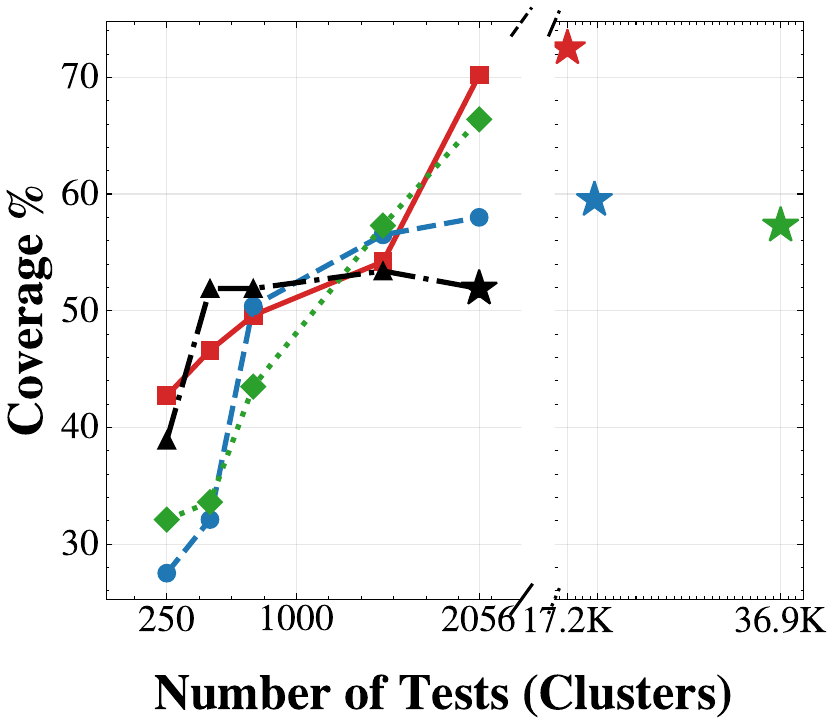}
        \caption{App Description}
        \label{fig:desc}
    \end{subfigure}\hfill
    \begin{subfigure}{0.27\textwidth}
        \includegraphics[width=\linewidth]{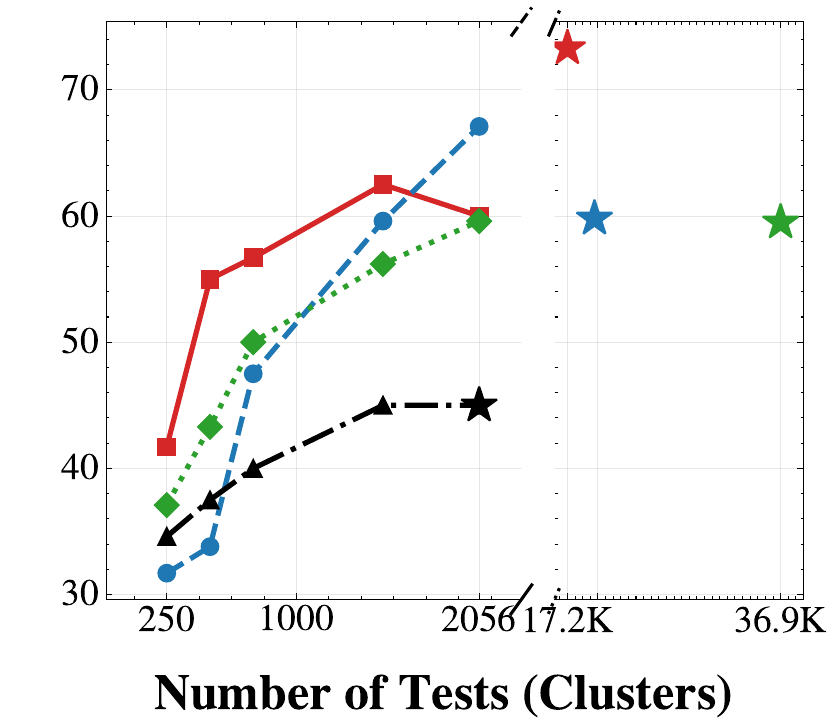}
        \caption{Positive Reviews}
        \label{fig:pos}
    \end{subfigure}\hfill
    \begin{subfigure}{0.27\textwidth}
        \includegraphics[width=\linewidth]{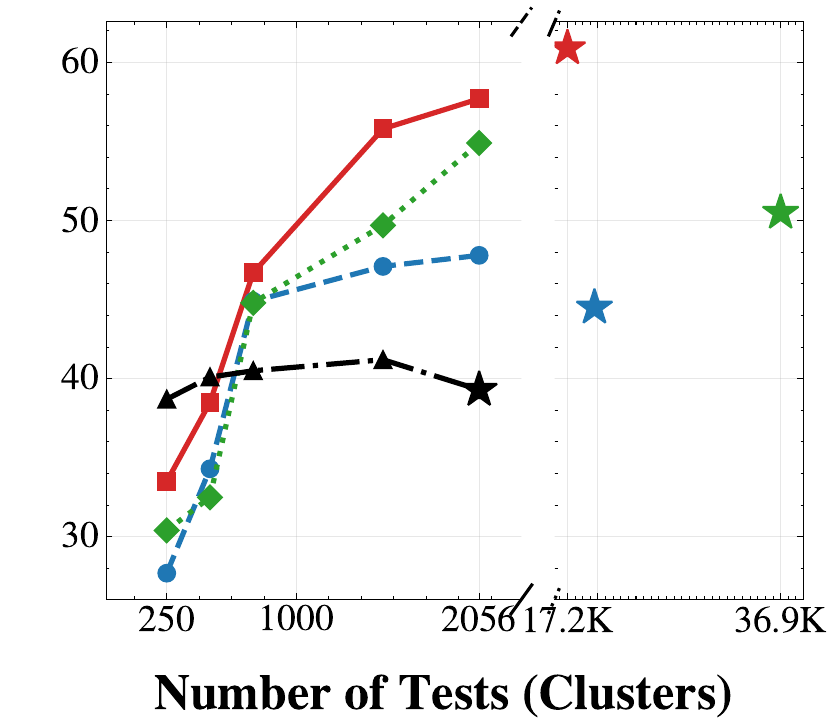}
        \caption{Negative Reviews}
        \label{fig:neg}
    \end{subfigure}
 
    \vspace{0.1cm}
 
    \includegraphics[width=0.5\textwidth]{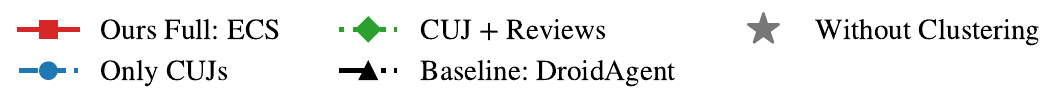}
 
    \caption{Detailed breakdown of coverage. Our simulation of real users through synthetic reviews substantially improves feature coverage.}
    \label{fig:breakdown}
    \vspace{-10pt}
\end{figure*}
 
 \vspace{-10pt}
\section{Results and Discussion}
 \vspace{-5pt}
 
\subsection{Feature Coverage Results}
 \vspace{-5pt}

\begin{wraptable}{r}{0.55\textwidth}
\centering
\vspace{-20pt}
\caption{Feature coverage at matched ($K=2056$) and full number of tests (no clustering).}
\label{tab:main_results}
\resizebox{0.53\textwidth}{!}{%
\begin{tabular}{l|c|cccc}
\toprule
\textbf{Method} & \textbf{K} & \textbf{Desc\%} & \textbf{Pos\%} & \textbf{Neg\%} & \textbf{Tot\%} \\
\midrule
DroidAgent (Full) & 2056 & 51.9\% & 45.0\% & 39.3\% & 40.3\% \\
\midrule
\textbf{Ours: ECS (Matched)} & 2056 & 70.2\% & 60.0\% & 57.7\% & 58.0\% \\
\quad Only CUJs & 2056 & 58.0\% & 67.1\% & 47.8\% & 51.8\% \\
\quad Ensemble: CUJs + Reviews & 2056 & 66.4\% & 59.6\% & 54.9\% & 55.6\% \\
\midrule
\textbf{Ours: ECS (Full)} & 17188 & \textbf{72.5\%} & \textbf{73.3\%} & \textbf{60.9\%} & \textbf{64.1\%} \\
\quad Only CUJs (Full) & 19682 & 59.5\% & 59.8\% & 44.5\% & 46.9\% \\
\quad Ensemble (Full) & 36870 & 57.3\% & 59.5\% & 50.5\% & 51.8\% \\
\bottomrule
\end{tabular}%
}
\vspace{-10pt}
\end{wraptable}

An important aim for testing frameworks is to create reproducible test cases that can be repeatedly used to comprehensively test an application. To achieve this, there is a trade-off between running every generated test and selecting a representative test suite, which is why we use clustering. Our main results across varying numbers of clusters are shown in Figure~\ref{fig:breakdown} and Table~\ref{tab:main_results}. We define \emph{Only CUJs} and the \emph{Ensemble: CUJs + Reviews} as ablations to isolate the impact of our synthetic review generation, allowing us to directly compare single CUJs against the combinations of CUJs used in our full \emph{ECS} method. The results diverge when broken down by source. \emph{Only CUJ} has high coverage of Positive Reviews, likely because positive reviews often mention features that are discrete and well captured by single interaction sequences. Conversely, our full method aggregates multiple CUJs into synthetic reviews and scores highest on both App Description (72.5\%) and Negative Review (60.9\%). Negative reviews frequently describe obscure edge cases or multi-step failures, which suggests that aggregating disjointed CUJs is an effective way to capture complex features. However, without the single CUJs, it is likely to miss the simpler core features. The \emph{Ensemble: CUJs + Reviews} ablation attempts to combine both but shows lower coverage compared to using only reviews, showing that focusing the generation budget to diverse synthetic reviews yields the best overall performance.

Covering 72.5\% of the application descriptions shows that ECS successfully captures the most prominent features. DroidAgent shows significantly lower coverage, likely because it infers test cases before visiting a screen, making it unlikely to discover complex features that require direct screen observation.

\textbf{Isolating the source of the gains.} To test whether the gain comes from visiting more screens, we filtered ECS to only the screens DroidAgent visited. Even with screens matched, ECS achieves 45.9\% versus DroidAgent's 40.3\%, and removing \texttt{missing\_gap} CUJs yields 46.3\%, showing each component contributes (Appendix~\ref{appendix:isolation}). We also report precision and F1 (Appendix~\ref{appendix:precision}) and evaluate GPTDroid and LLMDroid, which perform worse than both methods (Appendix~\ref{appendix:baseline_adapt}).

\textbf{Robustness across model families.} Our main results use Gemini models for generation, embedding, and judging, which raises the concern that models from the same family could generate and judge in mutually reinforcing ways~\citep{wang2024large}. We re-ran the judge with GPT-5.4-mini and Gemini~3.1~Pro, and separately re-ran the full CUJ and review generation with GPT-5.4-mini. ECS outperformed DroidAgent in every configuration (Appendix~\ref{appendix:robustness}). We also checked that the gains do not come from memorization of Play Store reviews. Using Gemini's training cutoff (December 17, 2025), the coverage gap between pre-cutoff and post-cutoff targets is only 3.8pp (Appendix~\ref{appendix:memorization}).


\vspace{-10pt}
\subsection{Ablations: Persona, Context Composition, and Number of Synthetic Reviews}
\vspace{-5pt}
\label{sec:ablations}
 
We study the different effects of our design choices through controlled ablations. For the ablations, we continue to use the DroidAgent's number of tests of $K=2056$.
 
\begin{wraptable}{r}{0.5\textwidth}
\vspace{-13pt}
\centering
\caption{Persona type and few-shot (FS) review prompting effects on feature coverage ($K=2056$).}
\label{tab:persona}
\scriptsize
\setlength{\tabcolsep}{3pt}
\begin{tabular}{llrrrr}
\toprule
\textbf{Persona} & \textbf{Prompt} & \textbf{Desc \%} & \textbf{Pos \%} & \textbf{Neg \%} & \textbf{Tot \%} \\
\midrule
No Persona & No FS & 53.4\% & 54.6\% & 49.3\% & 50.1\% \\
No Persona & FS    & 55.7\% & 55.0\% & 53.4\% & 54.2\% \\
\midrule
Lenient    & No FS & 43.5\% & 47.9\% & 48.7\% & 49.0\% \\
Lenient    & FS    & 58.0\% & 59.2\% & 53.7\% & 54.7\% \\
\midrule
Angry      & No FS & 61.1\% & 62.5\% & 53.2\% & 55.5\% \\
Angry      & FS    &62.6\% & 55.0\% & 51.3\% & 51.4\% \\
\midrule
Random       & No FS & 47.3\% & 56.7\% & 53.7\% & 54.2\% \\
\textbf{Random} & \textbf{FS [ECS]} & 70.2\% & 60.0\% & 57.7\% & \textbf{58.0\%} \\
\bottomrule
\end{tabular}
\vspace{-5pt}
\end{wraptable}
 
\textbf{Persona type and few-shot reviews.}
Table~\ref{tab:persona} studies the effect of persona type and few-shot prompting format. A few-shot prompt provides around 250 review examples drawn from a random, different application in our benchmark. These are the same examples provided to create the 40 to 50 user personas. While adding few-shot examples without a persona yields an improvement, particularly for reviews, it is less effective than explicitly building diverse personas. Comparing No Persona (FS) against Random (FS) shows a significant gap in App Description coverage (55.7\% \textit{vs} 70.2\%). The Random persona achieves this by sampling from a pool of 40 to 50 user personas. This indicates that injecting diverse user perspectives is helpful when the model is aggregating batches of CUJs, and is further improved when few-shot examples are provided. Fixed emotional personas, such as an exclusively angry or lenient user, do not improve performance.

\begin{wraptable}{r}{0.5\textwidth}
\vspace{-13pt}
\centering
\caption{Context composition showing the benefit of cross-screen CUJ sampling.}
\label{tab:composition}
\scriptsize
\setlength{\tabcolsep}{4pt}
\begin{tabular}{lrrrr}
\toprule
\textbf{Composition} & \textbf{Desc \%} & \textbf{Pos \%} & \textbf{Neg \%} & \textbf{Tot \%} \\
\midrule
16 Act, 0 Rand  & 45.0\% & 53.3\% & 52.5\% & 51.1\% \\
4 Act, 4 Rand   & 45.8\% & 49.2\% & 48.2\% & 47.8\% \\
8 Act, 8 Rand   & 54.2\% & 51.2\% & 46.3\% & 47.2\% \\
12 Act, 12 Rand & 45.8\% & 57.1\% & 49.6\% & 50.4\% \\
\textbf{16 Rand [ECS]} & 70.2\% & 60.0\% & 57.7\% & \textbf{58.0\%} \\
\bottomrule
\end{tabular}
\vspace{-5pt}
\end{wraptable}
 
\textbf{Context composition.}
Table~\ref{tab:composition} studies the effect of context composition. When all 16 CUJs are drawn from the same application screen (16 Act, 0 Rand), the model achieves low App Description coverage. This suggests that combining CUJs from similar screens is not very helpful for discovering new behaviors. Mixing activity and random CUJs in equal proportions (8 Act, 8 Rand) marginally improves app description coverage, but decreases overall coverage. We observed the highest performance when sampling CUJs uniformly at random from across the application (16 Rand). This is consistent with our persona findings, that diversity in context, like diversity in persona, is an important factor for aggregating CUJs effectively.

\begin{wraptable}{r}{0.5\textwidth}
\vspace{-15pt}
\centering
\caption{Context window size. Both too few and too many CUJs degrade performance.}
\label{tab:window_size}
\scriptsize
\setlength{\tabcolsep}{4pt}
\begin{tabular}{lrrrr}
\toprule
\textbf{Sample Size} & \textbf{Desc \%} & \textbf{Pos \%} & \textbf{Neg \%} & \textbf{Tot \%} \\
\midrule
1 CUJ  & 54.2\% & 49.2\% & 53.9\% & 52.9\% \\
8 CUJs & 53.4\% & 59.6\% & 53.7\% & 53.9\% \\
\textbf{16 CUJs [ECS]} & 70.2\% & 60.0\% & 57.7\% & \textbf{58.0\%} \\
32 CUJs & 45.0\% & 57.5\% & 52.2\% & 52.2\% \\
64 CUJs & 38.2\% & 49.2\% & 48.8\% & 48.4\% \\
\bottomrule
\end{tabular}
\vspace{-8pt}
\end{wraptable}
 
\textbf{Context window size.}
Table~\ref{tab:window_size} shows how the number of randomly sampled CUJs affects coverage. With 1 to 8 CUJs, the model has limited structural context and produces less specific outputs. Performance peaks at 16 CUJs, but declines at 32 and 64 CUJs. A possible reason for this decline is that providing too many disjointed CUJs forces the LLM to generate disparate, random reviews that lose coherence. This suggests an optimal window size exists for aggregating context without overwhelming the model.
 
\begin{wraptable}{r}{0.5\textwidth}
\vspace{-10pt}
\centering
\caption{Number of reviews. Coverage peaks at 400 batches and drops at 800 possibly due to redundancy.}
\label{tab:budget}
\scriptsize
\setlength{\tabcolsep}{4pt}
\begin{tabular}{lrrrr}
\toprule
\textbf{Batches} & \textbf{Desc \%} & \textbf{Pos \%} & \textbf{Neg \%} & \textbf{Tot \%} \\
\midrule
100 samples & 54.9\% & 53.9\% & 54.4\% & 53.4\% \\
200 samples & 60.2\% & 57.3\% & 52.7\% & 52.4\% \\
\textbf{400 [ECS]} & 70.2\% & 60.0\% & 57.7\% & \textbf{58.0\%} \\
800 samples & 64.6\% & 53.9\% & 55.6\% & 55.8\% \\
\bottomrule
\end{tabular}
\vspace{-5pt}
\end{wraptable}
 
\textbf{Number of synthetic reviews.}
As we scale the number of reviews from 100 to 400 batches, feature coverage improves with more diverse user experiences (Table~\ref{tab:budget}). However, we observe a drop in performance when increasing the number of reviews to 800. It is possible that forcing the LLM to generate too many batches results in highly redundant reviews. During the clustering phase, this excess redundancy may crowd out more diverse features. This suggests that an optimal number of reviews exists to balance diversity and redundancy.

\begin{table}[h]
\centering
\scriptsize 
\caption{Activity (screens) and Fragment (sub-screens) Coverage. Small Apps are from the Themis benchmark~\citep{su2021benchmarking} and Large Apps are from the Play Store.}
\label{tab:exploration}
\begin{tabular}{l|cc|cc}
\toprule
\multirow{2}{*}{\textbf{Method}} & \multicolumn{2}{c|}{\textbf{Activity Coverage}} & \multicolumn{2}{c}{\textbf{Fragments Discovered}} \\
& \textbf{Small Apps} & \textbf{Large Apps} & \textbf{Small Apps} & \textbf{Large Apps} \\
\midrule
DroidBot   & 37.8\% & 3.4\%  & 243 & 471  \\
GPTDroid   & 33.3\% & 5.2\%  & 344 & 286  \\
DroidAgent & 65.9\% & 14.9\% & 178 & 1514 \\
\midrule
\textbf{Ours (Full)} & \textbf{70.4\%} & \textbf{39.1\%} & \textbf{362} & \textbf{2945} \\
\midrule
W/out static analysis  & 55.6\% \textcolor{red}{\scriptsize[$-14.8$\%]} & 25.0\% \textcolor{red}{\scriptsize[$-14.1$\%]} & 266 & 2558 \\
W/out knowledge base   & 68.9\% \textcolor{red}{\scriptsize[$-1.5$\%]}  & 34.5\% \textcolor{red}{\scriptsize[$-4.6$\%]}  & 318 & 2873 \\
W/out forced exploration  & 64.4\% \textcolor{red}{\scriptsize[$-6.0$\%]}  & 20.4\% \textcolor{red}{\scriptsize[$-18.7$\%]} & 285 & 1086 \\
\bottomrule
\end{tabular}
\vspace{-5pt}
\end{table}

\vspace{-5pt}
\subsection{Activity Coverage}
 \vspace{-5pt}

The quality of semantic generation is ultimately bounded by the quality and depth of structural exploration. Table~\ref{tab:exploration} shows that static code analysis, our knowledge base, and our forced exploration to all activities significantly improves both Activity (screens) and Fragment (sub-screens) coverage. Our method is able to produce more diverse tests because it reaches more unique application activities and fragments. 
We find that performing static analysis to build a better exploration prior (\textit{W/out static analysis}) and implementing forced exploration (\textit{W/out forced exploration}) are the most critical components for achieving high activity coverage. Removing forced exploration causes a significant drop in performance, particularly on large applications, as the agent fails to attempt navigation to many screens. This also explains the activity coverage gap with DroidAgent. Because DroidAgent must first infer a test case before navigating, it cannot guarantee visits to specific activities, and tends to revisit familiar screens rather than systematically exploring the full application. Our forced exploration directly addresses this by directing the agent to each known activity in turn. Analysis of cost is shown in Appendix~\ref{appendix:cost} and Figure~\ref{fig:token_cost_analysis}.

\vspace{-5pt}
\subsection{Failure Cases: Limitations of LLM Agents in Mobile Testing}
\vspace{-5pt}

\label{sec:failure}

  \begin{wraptable}{r}{0.5\textwidth}
\vspace{-8pt}
\centering
\caption{Functional domains consistently missed, categorized by underlying LLM limitation.}
\label{tab:failure_domains}
\scriptsize
\resizebox{\linewidth}{!}{
\begin{tabular}{lrl}
\toprule
\textbf{Functional Domain} & \textbf{Count} & \textbf{Example Missed Feature} \\
\midrule
E-Commerce \& Ads  & 259 & Paywall limits access \\
Media \& Streaming & 179 & Playback progress reset on re-launch \\
Productivity \& AI & 142 & Calendar event sharing with guests \\
Files \& Data Sync & 114 & Automated camera roll sync \\
Social \& Comm.    & 71  & Keyboard dismissal layout resizing \\
Identity \& Security & 53 & Session persistence across app restart \\
General            & 263 & Image preview pinch-to-zoom scaling \\
\bottomrule
\end{tabular}
}
\vspace{-5pt}
\end{wraptable}

Despite the improvements demonstrated by ECS, our system failed to discover 1,081 of 3,012 features (35.9\%). We found these failures could often be categorized as statelessness, and discoverability constraints from the structure of our exploration. These categories were created based on word matching as shown in Appendix~\ref{appendix:regex_failure}. Commonly missed functional domains are reported in Table~\ref{tab:failure_domains}.

\textbf{Statelessness.} The most common failure came from the stateless nature of LLM agents. Each exploration trace is processed as an independent session, with no persistent state accumulating across sessions. This makes features that require state verification difficult to capture. Features such as ``playback progress resets when the app is closed and reopened'' (Media \& Streaming, 179 cases), ``camera roll changes are not reflected in cloud backup without a manual sync'' (Files \& Data Sync, 114 cases), or ``session tokens expire unexpectedly'' (Identity \& Security, 53 cases) require an agent to perform an action, exit the application, re-enter, and compare the current state against a prior one. The solution likely requires persistent state representations across sessions~\citep{park_generative}.

A related failure is the tolerance of LLMs toward monetization (259 features), which was the largest missed domain. While a human reviewer experiencing an aggressive subscription dialog would reflect that in their review, an LLM agent tends to dismiss the dialog as normal navigation.

\textbf{Discoverability constraints.} A second failure pattern came from the structure of our exploration strategy. Because our agent navigates to each activity in turn and enumerates available UI elements within it, features that only emerge from transitions \emph{across} activities are systematically undersampled. Calendar event sharing with guests (Productivity \& AI, 142 cases), for example, requires creating an event in one context, adding a guest, saving, and observing the downstream effect from a different screen. Our exploration structures each of these screens independently but does not produce the ordered sequence that makes the cross-screen dependency visible. Our batching of CUJs through synthetic reviews could not guarantee that these features were covered.
 
A related failure includes gesture-based interactions such as pinch-to-zoom and swipe-to-delete. Our agent's action vocabulary of taps and typed inputs does not produce multi-finger gestures unless explicitly prompted. Additionally, the agent lacks reliable feedback about whether the gesture registered correctly on the rendered UI.
 

\textbf{Broader relevance beyond mobile testing.}
While our testbed is mobile applications, both the translate-retrieve-judge evaluation pipeline and the persona-augmented generation approach are domain-general and could be applied to any setting with large amounts of user feedback, such as web applications or APIs as we discuss further in Appendix~\ref{appendix:broader}.

\textbf{Limitations and Future Work.}
While ECS demonstrates significant improvements in user-centric feature coverage, our study has several limitations. First, although we validated our automated LLM-as-a-judge pipeline against a human-judged subset, relying on language models for evaluation can introduce subtle verbosity or self-preference biases~\citep{wang2024large}. Second, our failure categorization relies on keyword matching, which could misclassify features that use synonymous or domain-specific terminology. These category counts should therefore be interpreted as indicative. In addition, our manual filtering of 5,779 candidates to 3,012 targets used the explicit taxonomy in Table~\ref{tab:filter_taxonomy} with conservative exclusion criteria, but we did not measure formal inter-annotator agreement on this construction step. Third, our User-Centric Feature Coverage benchmark is currently bounded to 10 popular applications and a collection of 5,000 real user reviews. Real-world applications evolve rapidly, and features present in older app versions may no longer exist in the APK we test against, creating mismatches between what the agent observes and what the metric expects. Future work should scale this dataset, include version-specific review tracking, and build on stronger GUI agent frameworks with persistent memory across sessions.

\vspace{-15pt}
\section{Conclusion}
\vspace{-10pt}

In this work, we introduced the User-Centric Feature Coverage benchmark, which consists of 3,012 target features crawled from real user reviews and app descriptions, alongside an automated LLM-as-a-judge evaluation pipeline. We then proposed ECS, a method that translates exploration traces into Critical User Journeys (CUJs) and aggregates them using persona-augmented synthetic reviews to capture complex features and write reproducible test cases. ECS achieves 64.1\% overall feature coverage, representing a 23.8\% gain over the state-of-the-art baseline, and covers 72.5\% of target app descriptions, capturing the most prominent features. Finally, our failure analysis reveals that LLM agents consistently struggle with stateful behaviors requiring cross-session reasoning and discoverability. These failures suggest that persistent episodic memory and cross-activity reasoning are important directions for future mobile testing frameworks.

\bibliography{colm2026_conference}
\bibliographystyle{colm2026_conference}

\newpage

\appendix

\section{Extended Related Work}
\label{appendix:related_extended}
\textbf{Mining the Play Store for Features.}
Several studies demonstrate that user reviews are a reliable source of features. Studies such as T-FREX~\citep{motger2024t} showed that review-derived feature sets capture features that static documentation frequently omits, and that many reviews implicitly mention feature requests~\citep{maalej2015bug,pagano2013user}. App description mining has similarly been used to generate test oracles and identify intended behaviors~\citep{gorla2014checking,shamsujjoha2021checking}. These studies support the idea that public data is a validated proxy for user-centric features.

\textbf{LLMs as Evaluators.}
LLM-as-a-Judge~\citep{gu2024survey, zheng2023judging} has become a standard evaluation paradigm in many NLP applications, with LLMs used as proxies for human preference across various tasks like open-ended generation, instruction following, and particularly for question answering~\citep{es2024ragas}. However, known risks include verbosity bias and self-preference bias~\citep{wang2024large}. We mitigate these risks by first translating all text into brief, objective description of the feature being tested.

\section{Identifying Missing Features}
\label{appendix:missing}
While our primary evaluation focuses on the features successfully tested by the agent, one of the most valuable secondary features of ECS is its ability to identify missing features. Because our framework explicitly compares expected features against the observed user interface during CUJ generation, it records functionalities that users expect but cannot find.

Table~\ref{tab:missing_features} shows a few examples from our Amazon dataset where real users complained about a missing, removed, or broken feature. Our system successfully captured the raw, frustrated user syntax, translated it into an objective target, and mapped it to a standardized system feature. Some negative-review targets describe functionality removed from current versions, such as ``Removal of the `save for later' feature.'' We argue these are among the most valuable targets, since they capture user-expected functionality that developers have removed or never implemented. A \texttt{missing\_gap} CUJ can only cover such a target when it matches the real user-reported feature, so these targets do not reward speculation (Appendix~\ref{appendix:isolation}).

\begin{table*}[h]
\centering
\caption{Examples of Missing Features Captured by ECS (Amazon Dataset). The system successfully translates diverse negative syntax into standardized system features.}
\label{tab:missing_features}
\scriptsize
\begin{tabular}{p{0.3\linewidth}p{0.3\linewidth}p{0.3\linewidth}}
\toprule
\textbf{Original User Review} & \textbf{Translated Target Feature} & \textbf{Matched Synthetic Feature} \\
\midrule
"Missing 'Prime' filter button." & Prime-specific search filtering. & Filter search results by Prime eligibility \\
"Ability to add items to a wishlist has disappeared from the app." & Wishlist 'Add to List' functionality. & Save item to a Wishlist from search results \\
"Removal of the 'save for later' feature." & 'Save for later' cart functionality. & View 'Saved for Later' items directly in Cart \\
"Missing 'sort by price' button." & Product sorting by price. & Search result sorting by price. \\
"No way to send bug reports directly to Amazon." & In-app bug reporting and feedback mechanism. & In-app feedback and bug reporting interface. \\
\bottomrule
\end{tabular}
\end{table*}

\section{Agent Actions and State Representation}
\label{appendix:agent_actions}
Our mobile testing agent interacts with the target application using a standard set of simple control tools alongside a custom state verification function. The action space includes basic gestures such as swipe, click, long click, and opening the application. To comprehend the application state, the agent parses the Android accessibility tree (a11y tree), which provides a structured text representation of all visible user interface elements. Furthermore, the agent utilizes a specialized 'check current screen' tool. This tool allows the agent to explicitly verify the current activity and sub-screen fragment it is navigating, ensuring that the LLM maintains an accurate understanding of its location within the application hierarchy during forced exploration.

\section{Isolating the Source of Coverage Gains}
\label{appendix:isolation}
ECS differs from DroidAgent in both exploration (static analysis, a knowledge base, and forced exploration) and test generation (CUJs and synthetic reviews). This raises the question of how much of the gain comes from just visiting more screens. To answer this, we took ECS trajectories and filtered them to only the CUJs and synthetic reviews derived from activities DroidAgent actually visited. We also regenerated the entire pipeline without \texttt{missing\_gap} prompts. Table~\ref{tab:isolation} reports feature coverage at matched $K{=}2056$.
 
Even with the screens matched, ECS outperformed DroidAgent (45.9\% \textit{vs} 40.3\%), showing that our CUJ generation and synthetic reviews contribute to performance on their own. At the same time, the gap to the full method (58.0\%) shows that improved exploration is itself an important contribution. The primary goal of testing is to test the app, and features that cannot be reached cannot be tested. Removing \texttt{missing\_gap} CUJs lowers coverage to 46.3\%, showing that they are also an important component.
 
We also note that \texttt{missing\_gap} generation cannot inflate the coverage score. A \texttt{missing\_gap} CUJ only counts toward coverage when it matches a real user-reported target, so candidates that do not correspond to real user-reported features are never counted. Prompting the model to imagine features is closer to the approach of DroidAgent, which asks the LLM to first infer a feature that `should' be present, then navigate to the screen that `should' support that feature. Our approach inverts this. We first verify all reachable screens, then infer which features are missing given what the LLM has actually observed.
 
\begin{table}[h]
\centering
\scriptsize
\caption{Isolating the contributions of exploration and generation (matched $K{=}2056$).}
\label{tab:isolation}
\begin{tabular}{lr}
\toprule
\textbf{Setup} & \textbf{Coverage (\%)} \\
\midrule
DroidAgent (own screens) & 40.3 \\
ECS masked to DroidAgent's screens & 45.9 \\
ECS without \texttt{missing\_gap} CUJs & 46.3 \\
\textbf{ECS (full, matched)} & \textbf{58.0} \\
\bottomrule
\end{tabular}
\end{table}

\section{Examples of Generated Test Cases}
\label{appendix:cuj_examples}
Table~\ref{tab:cuj_examples} provides concrete examples of the Critical User Journeys (CUJs) generated by our system for the Amazon Shopping application. Each CUJ pairs a high-level user intent with the exact, reproducible interaction steps and the underlying structural Activity/Fragment Path required to execute the test.

\begin{table*}[h]
\centering
\caption{Examples of generated Critical User Journeys (CUJs) translated into reproducible test cases.}
\label{tab:cuj_examples}
\resizebox{\linewidth}{!}{
\scriptsize
\begin{tabular}{p{0.15\linewidth}p{0.14\linewidth}p{0.21\linewidth}p{0.19\linewidth}p{0.14\linewidth}}
\toprule
\textbf{Test Name (CUJ)} & \textbf{User Goal} & \textbf{Test Steps} & \textbf{Activity/Fragment Path} & \textbf{Expected Result} \\
\midrule
Access Shopping Cart to complete an existing purchase & The user wants to check their cart status before deciding whether to continue shopping or sign in. & 
1. Action: Tap on the Cart icon. \newline Check: A Cart icon is visible in the header or footer. \newline\newline 
2. Action: Verify items in the cart. \newline Check: The shopping cart contents are displayed. & 
1. \texttt{MainActivity/} \newline \texttt{SigninPromptFragment/} & 
A persistent cart icon allows the user to see their pending items even when on a legal notice or onboarding screen. \\
\midrule
Search for Specific Product from Home Screen & Find a specific item to purchase. & 
1. Action: Tap on the global search bar at the top of the screen. \newline Check: The user is on the main gateway. \newline\newline 
2. Action: Type 'organic coffee' and press the search icon or enter key. \newline Check: The keyboard appears and text is entered. \newline\newline 
3. Action: Verify the search results page loads. \newline Check: Search results for organic coffee are shown. & 
1. \texttt{MainActivity/} \newline \texttt{GatewayMigration}\newline\texttt{Fragment/} \newline\newline 
2. \texttt{MainActivity/} \newline \texttt{GatewayMigration}\newline\texttt{Fragment/} \newline\newline 
3. \texttt{MainActivity/} \newline \texttt{ProductSearch}\newline\texttt{XRayFragment/}  &
The search results page for the entered query is displayed correctly. \\
\bottomrule
\end{tabular}
}
\end{table*}

\section{Persona and Role-Playing in Language Models}
\label{appendix:persona_related_work}
Prompting LLMs to adopt specific personas is a well-studied area in HCI and NLP~\citep{ha2024clochat, schuller2024generating}. CloChat~\citep{ha2024clochat} synthesizes 25 HCI studies on persona design and finds that attributes such as demographics, personality traits, and verbal style each contribute to perceived trust and empathy in user interactions. Furthermore, \cite{schuller2024generating} demonstrated that LLM-generated personas are rated as credible as human-written ones by UX researchers. These studies suggest that personas reliably shift model behavior in ways users find meaningful, largely because LLMs are highly capable at impersonating distinct viewpoints. In our work, we find that persona diversity specifically benefits feature coverage. By simulating diverse users, the model generates a much wider variety of feature interactions than it would using a single default perspective.

\section{Application Lists}
\label{appendix:app_list}
\textbf{Small Apps (Themis):} ActivityDiary, AnkiDroid, AntennaPod, Markor, Omni-Notes, Phonograph, Scarlet-Notes, Commons, Openlauncher, Osmeditor4android.
\textbf{Large Apps (Play Store):} Dropbox, edjing~mix, Audible, Amazon Shopping, BBCiPlayer~(UK), Discord, Calendar, Fotor, Sheets, Flipkart.

\section{Prompts}
\label{appendix:prompts}
This section details the exact prompts used across the three phases of our method: exploration, CUJ generation, synthetic review generation, as well as the prompts used for automated evaluation.

\subsection{Exploration Prompts}
During the exploration phase, the agent is provided with a target activity and a memory string. The `memory` variable contains the specific transition path and UI elements required to navigate to the target location. 

\begin{quote}
\small \ttfamily
MISSION:\\
1. **NAVIGATE**: Go to '\{current\_activity\}'.\\
\\
2. **KNOWLEDGE BASE**: \\
\{(Any known path and UI elements)\}\\
\\
3. **CORE FEATURE RULES**: \\
- If you see a *NEW* text input, type and send.\\
- If you see a list, click an item you haven't clicked before.\\
- If stuck, go BACK.\\
- Continue exploring. Look for different activities that can be activated within the target activity. Look for different fragments.\\
\\
RESTRICTIONS:\\
- STAY IN APP: No Settings/Play Store.\\
- When typing to boxes, provide waits after clicking each box. Wait at least 1 second after clicking a box before typing. Also be sure to clear the textbox before typing if needed.\\
- Max Steps: 15.\\
- If you change languages, make sure you change the language back to english before ending.\\
- Use print(get\_current\_activity\_path()) to check where you are.\\
- If you ever encounter an 'EMPTY' state, you must wait for at least 1 second and run the print(get\_current\_activity\_path()) again. This just means the UI has not loaded properly yet.
\end{quote}

\subsection{Test Generation Prompts}
To generate Critical User Journeys (CUJs), we first establish the LLM's role as a product visionary using a system prompt. This encourages the model to evaluate both implemented features and missing expectations. 

\begin{quote}
\small \ttfamily
You are an **Expert Product Manager and UX Visionary** analyzing an Android app's 'Exploration Trace'.\\
Your task is to generate an EXHAUSTIVE, comprehensive list of Critical User Journeys (CUJs) for the target screen.\\
\\
**CORE INSTRUCTIONS:**\\
\\
1. **Exhaustive UI Coverage (Implemented Features):** \\
Meticulously scan the provided Target Screen UI tree. You must generate a distinct CUJ for ALMOST EVERY actionable/touchable element (buttons, tabs, inputs, list items, menus). If a UI element exists, a user has a goal for it. Do not group them too broadly. Mark these as `implemented`.\\
\\
2. **World Knowledge \& The "Ideal App" (Missing Features):** \\
Use your domain expertise. What is this app trying to be? When a user lands on this specific screen, what do they *expect* to be able to do that isn't currently supported by the UI? Think about industry standards, edge cases, and power-user needs. Generate CUJs for these expected ideal features and mark them as `missing\_gap`.\\
\\
You must extract as many CUJs as needed to cover important interactive UI elements PLUS missing ideal features based on world knowledge.
\end{quote}

We then provide the trace and global history to extract the individual CUJs:

\begin{quote}
\small \ttfamily
\{trace\_formatted\}\\
\\
PREVIOUSLY GENERATED CUJS FOR THIS HUB (Do not repeat these exact goals unless focusing on a different UI element):\\
- \{global\_history\}\\
\\
TASK:\\
1. Read the Final Step UI carefully. Write CUJs for distinct touchable/interactive elements.\\
2. Access your world knowledge. Think of **IDEAL FEATURES.** What features SHOULD be on this screen (e.g. if it's a chatting app, it should probably have group chat options) but are missing? Write CUJs for them.\\
Return the exhaustive list.
\end{quote}

For generating persona-augmented synthetic reviews, we inject the generated persona string alongside batches of CUJs:

\begin{quote}
\small \ttfamily
\{current\_persona\_str\}\\
\\
--- EXPLORATION TRACES ACCESSED ---\\
\{combined\_traces\}\\
\\
--- FEATURES ENCOUNTERED \& MISSING ---\\
\{cuj\_summary\}\\
\\
--- PREVIOUSLY GENERATED REVIEWS (DO NOT REPEAT THESE) ---\\
\{history\_str\}\\
\\
TASK: Write MULTIPLE highly specific, dense reviews. Ensure they are DIFFERENT from the previously generated reviews above.
\end{quote}

\subsection{Automated Evaluation Prompts}
To evaluate feature coverage, we first translate both synthetic reviews and target features into objective functionality descriptions.

\begin{quote}
\small \ttfamily
You are a QA Engineer. Read the following and translate the underlying app mechanics into OBJECTIVE, TECHNICAL feature descriptions. Do NOT include user opinions, bugs, or emotions. Identify the core functional feature being described or tested.\\
\\
Example 1 (Review): 'The stupid cart won't let me delete items!' -> 'Shopping cart item deletion functionality.'\\
Example 2 (Test Case): 'Click the profile avatar then click dark mode toggle' -> 'Dark mode theme toggle.'\\
\\
Extract the objective features mapped to Index:
\end{quote}

Finally, the LLM-as-a-judge makes a binary decision using the translated texts:

\begin{quote}
\small \ttfamily
Review these items. Determine if ANY candidate test strongly covers the Feature. \\
Candidates: \{5\_candidates\} \\
Target Feature: \{1\_target\_feature\}
\end{quote}

\section{Additional Baselines: GPTDroid and LLMDroid}
\label{appendix:baseline_adapt}
GPTDroid and LLMDroid optimize for Activity Coverage and produce step-level textual logs rather than natural language test cases. To evaluate them for user-centric feature coverage, we applied a translation of their outputs into user goals before passing them to the judge, using the same translation prompt as all other methods. We evaluated at matched $K{=}2056$. As shown in Table~\ref{tab:steplevel}, LLMDroid and GPTDroid underperform compared to ECS and DroidAgent. This is most likely a combination of lower activity coverage (Table~\ref{tab:exploration}) and the absence of discrete exploration traces, which would translate more cleanly into user goals.
 
\begin{table}[h]
\centering
\scriptsize
\caption{Feature coverage of step-level baselines translated into user goals (matched $K{=}2056$).}
\label{tab:steplevel}
\begin{tabular}{lrr}
\toprule
\textbf{Method} & \textbf{K} & \textbf{Coverage (\%)} \\
\midrule
\textbf{ECS (Matched)} & 2056 & \textbf{58.0} \\
DroidAgent & 2056 & 40.3 \\
LLMDroid (Matched) & 2056 & 25.6 \\
GPTDroid (Matched) & 2056 & 24.7 \\
\bottomrule
\end{tabular}
\end{table}

\section{Precision, Recall, and F1 of Generated Tests}
\label{appendix:precision}
Feature coverage measures how many benchmark targets are covered, which is analogous to recall. It does not measure what fraction of the generated tests is effective. We therefore compute precision as the fraction of test clusters whose top-5 retrieved targets include at least one pair judged Pass, over the total number of clusters. Table~\ref{tab:pr} reports precision, recall, and F1 at different values of $K$. ECS at $K{=}500$ matches DroidAgent's peak F1 while reaching a substantially higher recall ceiling. Precision at high $K$ naturally drops for recall-maximizing methods, since additional clusters increasingly capture redundant behaviors.
 
\begin{table}[h]
\centering
\scriptsize
\caption{Precision, recall, and F1 of generated test suites at different values of $K$.}
\label{tab:pr}
\begin{tabular}{llccc}
\toprule
\textbf{Method} & \textbf{K} & \textbf{Precision} & \textbf{Recall} & \textbf{F1} \\
\midrule
ECS & 500 & 0.63 & 0.43 & 0.51 (max F1) \\
ECS & 2056 & 0.35 & 0.58 & 0.44 \\
DroidAgent & 250 & 0.75 & 0.38 & 0.51 (max F1) \\
DroidAgent & 2056 & 0.27 & 0.40 & 0.33 \\
\bottomrule
\end{tabular}
\end{table}

\section{Robustness Across Model Families}
\label{appendix:robustness}
Our main results use Gemini models for generation, embedding, and judging. This raises the concern that models from the same family could generate and judge in mutually reinforcing ways~\citep{wang2024large}. We address this from two directions at matched $K{=}2056$ (Table~\ref{tab:robustness}). First, we re-ran the judge with GPT-5.4-mini and Gemini~3.1~Pro while keeping the generated tests fixed. Second, we re-ran the full CUJ and synthetic review generation with GPT-5.4-mini, using \texttt{text-embedding-3-small} (OpenAI's embeddings) for clustering. ECS outperformed DroidAgent in every configuration, showing that the benefit of ECS is not limited to Gemini models.
 
\begin{table}[h]
\centering
\scriptsize
\caption{Feature coverage (\%) at matched $K{=}2056$ with different judge and generation models. Left: tests generated with Gemini~3~Flash and evaluated by different judges (\texttt{gemini-3-flash-preview}, \texttt{gpt-5.4-mini-2026-03-17}, \texttt{gemini-3.1-pro-preview}). Right: the full generation pipeline re-run with GPT-5.4-mini, using \texttt{text-embedding-3-small} for clustering.}
\label{tab:robustness}
\setlength{\tabcolsep}{4pt}
\begin{tabular}{l|ccc|cc}
\toprule
& \multicolumn{3}{c|}{\textbf{Judge (Gemini-generated tests)}} & \multicolumn{2}{c}{\textbf{GPT-generated tests}} \\
\textbf{Method} & \textbf{Gemini 3 Flash} & \textbf{GPT-5.4-mini} & \textbf{Gemini 3.1 Pro} & \textbf{Gemini judge} & \textbf{GPT judge} \\
\midrule
\textbf{ECS (Matched)} & \textbf{58.0} & \textbf{61.3} & \textbf{46.8} & \textbf{44.2} & \textbf{49.9} \\
DroidAgent & 40.3 & 43.8 & 39.3 & 32.7 & 37.1 \\
\bottomrule
\end{tabular}%
\end{table}

\section{Checking for Memorization of Target Reviews}
\label{appendix:memorization}
Because our benchmark is built from publicly available Play Store reviews, reviews of popular applications may appear in LLM training data. We checked that the gains do not come from memorization of the target reviews. Using Gemini's training cutoff (December 17, 2025) and matched $K{=}2056$, we found that 34.8\% of review targets are post-cutoff and cannot be memorized. As shown in Table~\ref{tab:memorization}, the gap between pre-cutoff and post-cutoff coverage is only 3.8pp, showing that coverage mostly reflects observed app behavior rather than memorized reviews. App description targets are pre-existing and are covered at a higher rate, which is consistent with them representing the most prominent features.
 
\begin{table}[h]
\centering
\scriptsize
\caption{ECS coverage (matched $K{=}2056$) split by Gemini's training cutoff (December 17, 2025).}
\label{tab:memorization}
\begin{tabular}{lrr}
\toprule
\textbf{Target category} & \textbf{Total} & \textbf{Coverage} \\
\midrule
App descriptions (pre-existing) & 131 & 70.2\% \\
Reviews pre-cutoff & 1,879 & 58.8\% \\
Reviews post-cutoff (cannot be memorized) & 1,002 & 55.0\% \\
\midrule
Overall & 3,012 & 58.0\% \\
\bottomrule
\end{tabular}
\end{table}

\section{Cost analysis}
\label{appendix:cost}
 
\begin{figure*}[h]
    \centering
    \begin{subfigure}{0.3\textwidth}
        \includegraphics[width=\linewidth]{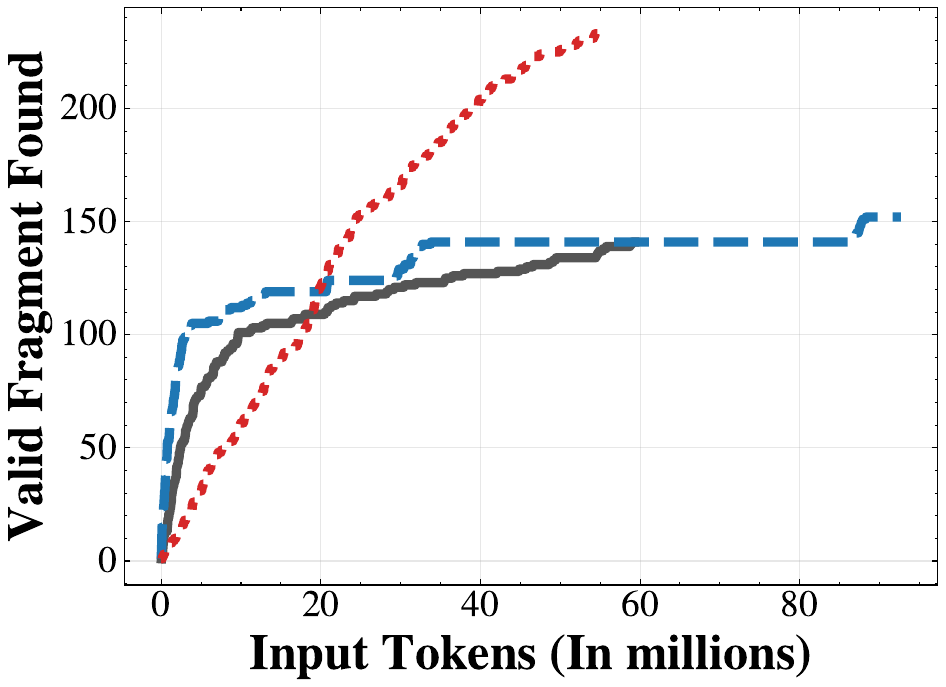}
        \caption{Input token usage}
        \label{fig:input_tokens}
    \end{subfigure}\hfill
    \begin{subfigure}{0.3\textwidth}
        \includegraphics[width=\linewidth]{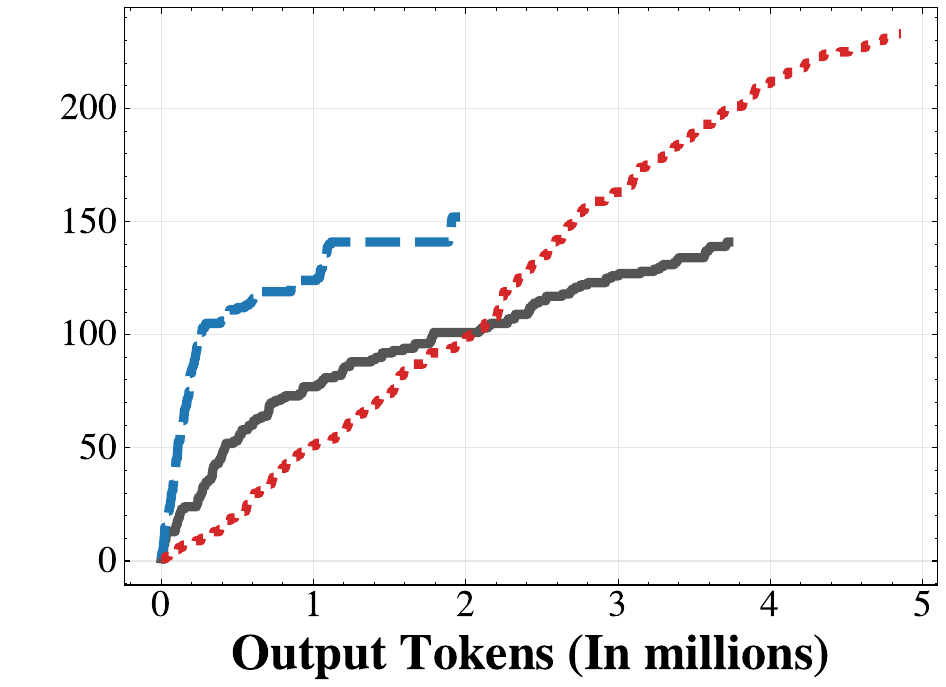}
        \caption{Output token usage}
        \label{fig:output_tokens}
    \end{subfigure}\hfill
    \begin{subfigure}{0.3\textwidth}
        \includegraphics[width=\linewidth]{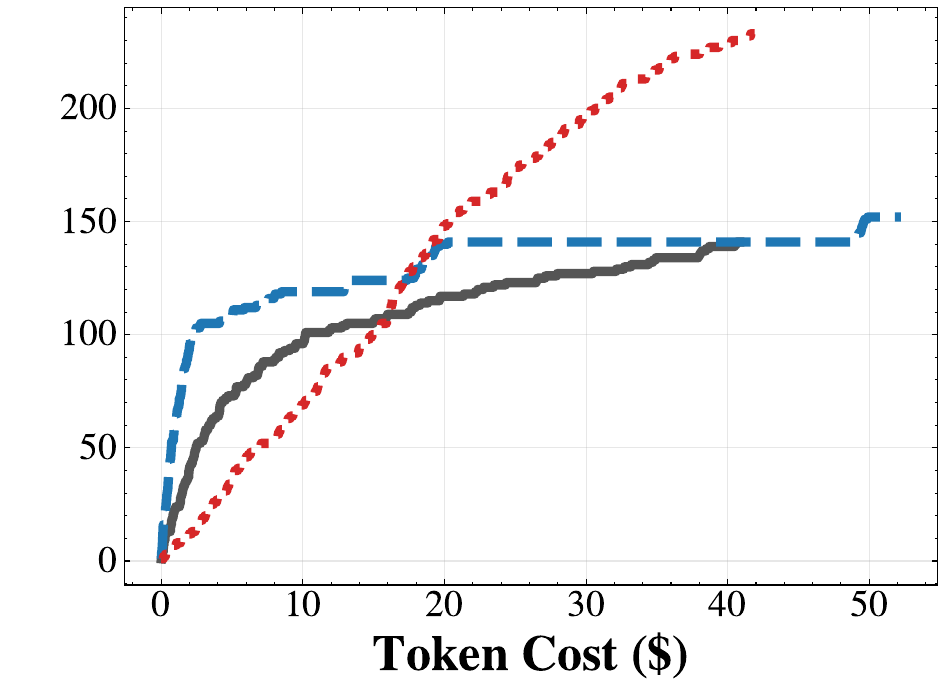}
        \caption{Total computational cost}
        \label{fig:cost_total}
    \end{subfigure}
    \vspace{0.2cm}
    \includegraphics[width=0.6\textwidth]{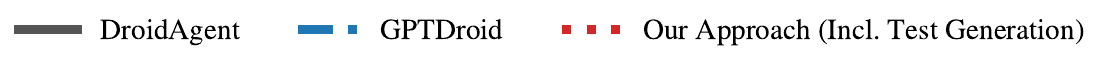}
    \caption{Token usage relative to unique fragments (sub-screens) discovered.}
    \label{fig:token_cost_analysis}
\end{figure*}

Figure~\ref{fig:token_cost_analysis} shows the cost of methods that use LLM calls. We find that our method is the most cost effective in the number of valid fragments found. This occurs because our LLM is not required to store a history of prior interactions. In contrast, baseline methods show less cost-effectiveness because they append previous instruction memory at each step. As these token payloads increase, API response times also degrade.

\section{Failure Analysis Categorization}
\label{appendix:regex_failure}
To analyze and summarize the features our method failed to capture (Section~\ref{sec:failure}), we used a keyword-based categorization approach. We defined six domains based on common mobile application features and mapped them to specific keyword arrays. The translated target features of the missed tests were then matched against these arrays to isolate the most prominent areas where LLM agents struggle. The categories and their associated keywords are defined as follows:

\begin{itemize}
    \item \textbf{Media \& Streaming}: \texttt{video}, \texttt{audio}, \texttt{music}, \texttt{podcast}, \texttt{playback}, \texttt{radio}, \texttt{stream}, \texttt{player}, \texttt{album}, \texttt{track}, \texttt{song}, \texttt{mp3}, \texttt{bpm}, \texttt{vocal}, \texttt{pitch}, \texttt{equalizer}, \texttt{episode}, \texttt{chapter}, \texttt{acapella}, \texttt{play next}
    
    \item \textbf{E-Commerce \& Ads}: \texttt{cart}, \texttt{checkout}, \texttt{pay}, \texttt{billing}, \texttt{subscription}, \texttt{monetization}, \texttt{ad}, \texttt{ads}, \texttt{advertisement}, \texttt{price}, \texttt{wishlist}, \texttt{order}, \texttt{delivery}, \texttt{locker}, \texttt{gift card}, \texttt{credit}, \texttt{premium}, \texttt{trial}, \texttt{loyalty}, \texttt{refund}, \texttt{invoice}, \texttt{shipping}, \texttt{fee}
    
    \item \textbf{Files \& Data Sync}: \texttt{file}, \texttt{folder}, \texttt{sync}, \texttt{download}, \texttt{upload}, \texttt{storage}, \texttt{pdf}, \texttt{zip}, \texttt{rar}, \texttt{cloud}, \texttt{backup}, \texttt{export}, \texttt{import}, \texttt{drive}, \texttt{document}, \texttt{save}, \texttt{offline}
    
    \item \textbf{Social \& Communication}: \texttt{chat}, \texttt{message}, \texttt{call}, \texttt{voice}, \texttt{server}, \texttt{role}, \texttt{friend}, \texttt{sticker}, \texttt{emoji}, \texttt{reaction}, \texttt{communication}, \texttt{dm}, \texttt{conferencing}, \texttt{bot}, \texttt{avatar}, \texttt{gif}
    
    \item \textbf{Identity \& Security}: \texttt{login}, \texttt{password}, \texttt{auth}, \texttt{account}, \texttt{sso}, \texttt{verification}, \texttt{permission}, \texttt{logout}, \texttt{session}, \texttt{security}, \texttt{passkey}, \texttt{2fa}, \texttt{recovery}
    
    \item \textbf{Productivity \& AI}: \texttt{ai}, \texttt{formula}, \texttt{cell}, \texttt{sheet}, \texttt{calendar}, \texttt{schedule}, \texttt{event}, \texttt{task}, \texttt{reminder}, \texttt{predictive}, \texttt{smart}, \texttt{object removal}, \texttt{face swapping}, \texttt{formatting}, \texttt{chart}, \texttt{dropdown}
\end{itemize}

\section{Broader Relevance Beyond Mobile Testing}
\label{appendix:broader}
While our testbed is mobile applications, the challenges and solutions in this work extend beyond black-box app testing. Our evaluation pipeline translates system outputs and user-written expectations into a shared semantic space before judging coverage. This design can be applied to any setting with large amounts of user feedback, such as web applications or APIs. Our generation results also apply beyond this domain. Diverse personas improve review generation (Table~\ref{tab:persona}), which is applicable to other settings that deal with human reviews, such as recommendation. In addition, our synthetic reviews show that LLMs can take a set of independent CUJs and combine them into coherent multi-step narratives, linking unrelated observations into realistic workflows. This composition could generalize to other settings requiring narrative synthesis from observations, such as summarizing execution traces or API call logs.

\section*{Ethics Statement}
Our goal is to help developers better understand their users and build higher-quality applications, and the intent of this work is constructive and developer-facing. Our benchmark stores no personal user information. All targets are derived from publicly available Play Store reviews and app descriptions, which were already public at crawling time. Personas are generated from reviews of unrelated applications, preventing leakage of individual user identity into test generation for any given target app. ECS runs the application binary in an isolated test environment with a fresh install and no real user data, and the intended deployment is developer-side testing during app development. Finally, as we mention in Section~\ref{sec:failure}, identity and security features are hard to test reliably, and we believe credential-handling sequences should remain human-supervised.

\end{document}